\documentclass[%
 aip,
 amsmath,amssymb,
preprint,%
 reprint,%
]{revtex4-2}

\usepackage{graphicx}
\usepackage{dcolumn}
\usepackage{bm}

\usepackage[utf8]{inputenc}
\usepackage[T1]{fontenc}
\usepackage{mathptmx}
\usepackage{etoolbox}
\usepackage{eucal}
\DeclareMathAlphabet{\mathscr}{OT1}{pzc}{m}{it}

\usepackage{color,soul}

\graphicspath{{Figures/}}

\makeatletter
\def\@email#1#2{%
 \endgroup
 \patchcmd{\titleblock@produce}
  {\frontmatter@RRAPformat}
  {\frontmatter@RRAPformat{\produce@RRAP{*#1\href{mailto:#2}{#2}}}\frontmatter@RRAPformat}
  {}{}
}%
\makeatother

\usepackage{physics}

\usepackage{bm, bbm, upgreek}

\newcommand{\begeq}{\begin{equation}\begin{gathered}}
\newcommand{\eqend}{\end{gathered}\end{equation}}
\newcommand{\begal}{\begin{equation}\begin{aligned}}
\newcommand{\alend}{\end{aligned}\end{equation}}

\newcommand{\ten}[1]{\bm{#1}}
\newcommand{\p}{\partial}
\newcommand{\del}{\updelta}

\begin{document}

\preprint{AIP/123-QED}

\title{Deposition Rates in Thermal Laser Epitaxy: Simulation and Experiment}
\author{Thomas J. Smart}
\affiliation{Peter Grünberg Institute -9, Forschungszentrum Jülich GmbH, Jülich, Germany}
\author{Bilen Emek Abali}
\affiliation{Department of Materials Science and Engineering, Uppsala University, \\
	Ångströmlab, Box 35, 751 03, Uppsala, Sweden}%
\author{Hans Boschker}
\author{Wolfgang Braun}
 \affiliation{epiray GmbH, Heinrich-Otto-Str. 73, 73240, Wendlingen am Neckar, Germany}
\email[Corresponding Author: ]{wbraun@epiray.de}

\date{\today}

\begin{abstract}
The modeling of deposition rates in Thermal Laser Epitaxy (TLE) is essential for the accurate prediction of the evaporation process and for improved dynamic process control. We demonstrate excellent agreement between experimental data and a model based on a finite element simulation that describes the temperature distribution of an elemental source when irradiated with continuous wave laser radiation. The simulation strongly depends on the thermophysical constants of the material, data of which is lacking for many elements. Effective values for the parameters may be determined with precision by means of an unambiguous reference provided by the melting point of the material, which is directly observed during the experiments. TLE may therefore be used to study the high temperature thermophysical and optical properties of the elements.

\end{abstract}

\maketitle

\section{Introduction}
Shortly after the invention of the laser in 1960,\cite{Maiman_1960} it was quickly realised that the unique properties of the laser beam had a multitude of applications. In the proceeding years, these applications included welding,\cite{laserheating_prokhorov,Hashimoto1991}  cutting,\cite{lasercutting} metrology \cite{LaserMetrology,LaserReflector} and laser-based lidar technology.\cite{Collis70} The localised heating produced by a laser beam was also found to drive irradiated materials to extremely high temperatures.\cite{ambartsumyan1965, Lichtman1963,Ready1965}
These temperatures allowed for refractory materials with high melting points to be easily evaporated for the growth of thin films.\cite{Smith:65,Groh,Hass69,Sankur85,Dijkkamp1987PreparationOY}


Continuous-wave (CW) laser heating of free-standing material sources forms the backbone of a new deposition technique called Thermal Laser Epitaxy (TLE).\cite{braun}
Via steady state thermal heating of these sources by laser radiation, TLE generates atomic fluxes that are then deposited upon a substrate. This allows for the deposition of a wide array of elements and compounds from across the periodic table.\cite{smart,carbonTLE,AlOxTLE,RuTLE}
Significant deposition rates for refractory metals like W and Ta are obtained when source temperatures typically exceed 3000\,K.\cite{smart,TaTLE} 
At these temperatures, non-linear energy loss processes like radiation and evaporation become significant and consequentially cannot be easily solved by an analytical model.
Therefore, in order to simulate the evaporation of metals resulting from CW laser heating, adequate modeling of the physical processes is required.

Computational models of laser heating primarily focus on applications for laser welding,\cite{MACKWOOD2005} which account for the effects of the extreme laser power densities, such as a plume or plasma formation.\cite{Ready1963-Plume}
However, the laser power densities typical in TLE are far lower than those mostly used by laser welding applications,\cite{smart} thus a thermal model is appropriate.
Such models have been very successful at replicating experimental results including the temperature dependence of the thermophysical parameters.\cite{Gallais2021, Elhadj2014,Combis2012} 

In this paper, we demonstrate a practical method to determine the parameters in the temperature range of interest and thereby reduce the number of necessary parameters for an accurate simulation of the CW laser heating process.
We use these effective values to accurately predict steady state thermal laser evaporation and compare the calculated evaporation rates with experimentally measured evaporation rates.

The behavior of the model suggests that a self-consistent simulation of a large number of experiments with different elemental materials allows for the precise determination of the thermophysical parameters of most elements at ultra-high temperatures.
This allows for an accurate predictive control of the TLE process for the synthesis of epitaxial heterostructures from practically all elements in the periodic table.

\begin{figure}[!b]
	\centering
	\includegraphics[width =\linewidth]{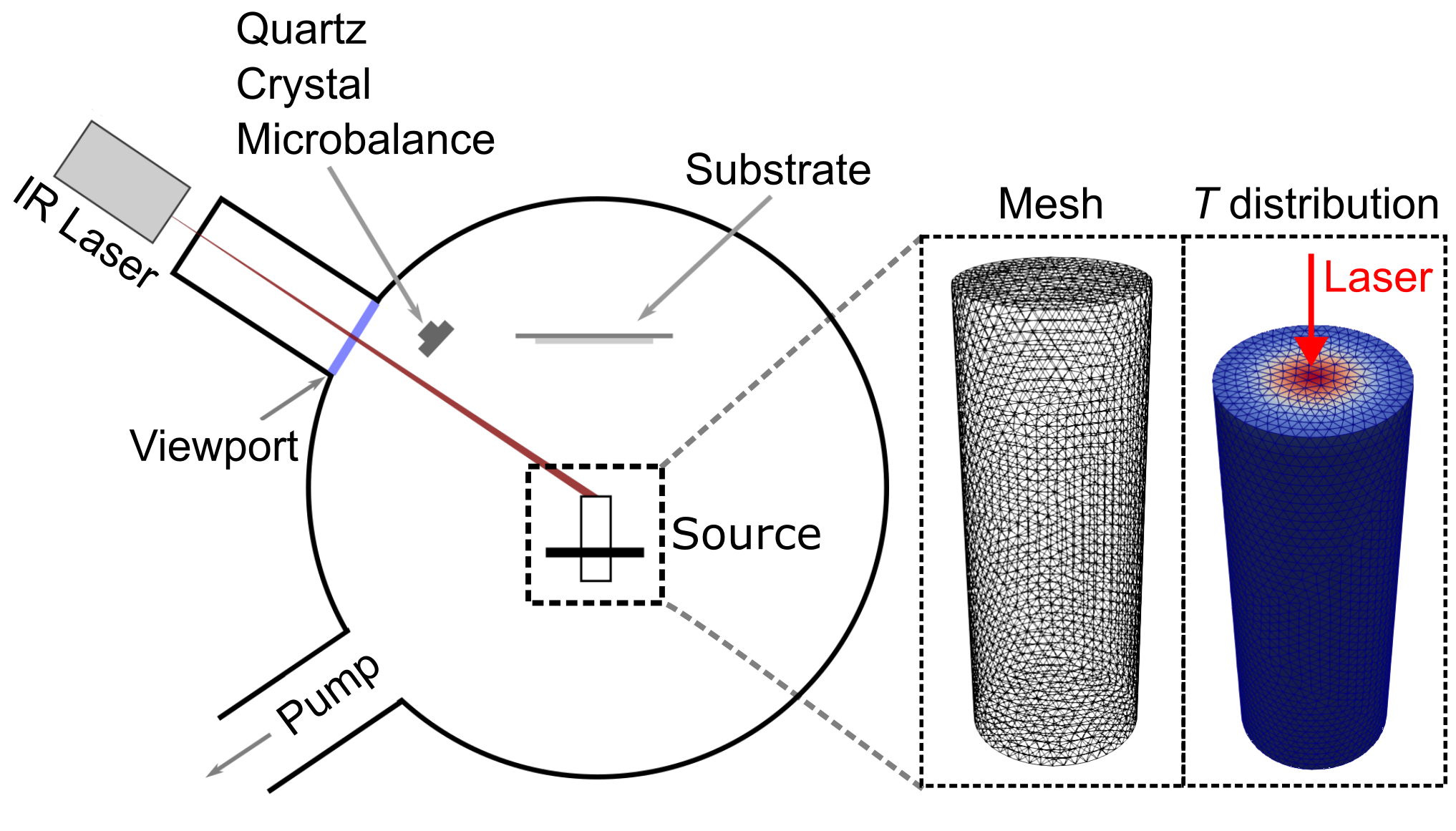}
	\caption{Cross-sectional sketch of the TLE chamber used for the measurements.
	The uncooled chamber was operated without substrate heating.
	The insets include a representation of the computational mesh of a source, along with an example of a resolved temperature distribution with a red arrow indicating the position of the incident laser.
	The geometry has a diameter of 3\,mm and a length of 8\,mm.}
	\label{Fig1}
\end{figure}

\section{Model}
\subsection{Mechanisms and terms}

Within this model, we aim to model the CW laser heating of a free-standing source of material as is performed in a simplified TLE chamber shown in Fig. 1. Within this system, the laser beam directly impinges on the surface of the cylindrical source, which is contained in a chamber operating at ultra-high vacuum conditions.\cite{braun} Previous work has demonstrated that during operation, the free-standing source exhibits strong temperature gradients due to the localized heating from the laser beam. Coupled with the minuscule contact surface between the source and the source holder, the temperature gradient results in minimal thermal conduction between the source and its surrounding environment.\cite{smart} Therefore, within the model, we consider the source as a cylindrical rod in a vacuum with the laser impinging on the upper surface, as demonstrated in the inset of Fig. 1.
\par
We begin by considering the internal energy balance for a rigid body under non-equilibrium thermodynamics:\cite{abali}
\begin{equation}
\rho \dot u + q_{i,i} - \rho r = 0 \ ,
\label{energybalance}
\end{equation}
where we use Einstein's summation convention over repeated indices, $i \in (1,2,3)$ in three-dimensional space with a comma notation for partial space derivatives and dot notation for rates.
$\rho$ denotes the mass density and $u$ is the specific internal energy in J/kg.
The (heat) flux term $q_i$, is defined as the flow of energy per unit time through a unit area of a surface.
The supply term $r$ represents the volumetric energy increase by a given flux of energy.
In the case of TLE, we model the incoming energy flux from a laser as being incident on the surface of the source. In other words, no bulk supply term is needed, so $r$ = 0.
Since we are dealing with a rigid body, the internal energy of the system is a function of the temperature $T$ only.
Here, the temperature rate is small such that the assertion that $u = u(T)$ is adequate without dependence on the temperature rate, which is the case in rational thermodynamics, otherwise one needs extended rational thermodynamics \cite{muller2013rational}. Hence, we define $u=u(T)$ and use a specific heat capacity $c = \p u / \p T$.
The distinction between isobaric and isochoric specific heat capacity can be neglected in this model as we assume that the material is rigid and does not deform.  
Moreover, we use Fourier's law as the simplest model for the flux term, as follows:
\begin{equation}
q_i = -\kappa T_{,i} ,
\label{fourierlaw}    
\end{equation}
where we model the material as isotropic.
Within this approach, we model the absorption of light from the incident laser beam on the irradiated surface of the source as a supply term $L(x,y)$, acting on the surface.
This supply term has a Gaussian profile and we assume that $L(x,y)$ is constant in time
\begin{equation}
    L(x,y) = \frac{(1-\mathcal{R})P}{\pi \omega^2}\exp\Big(\frac{x^2 + y^2}{\omega^2}\Big) \ ,
\label{lasersource}    
\end{equation}
where $\mathcal{R}$ is the reflectivity of the source material at the laser wavelength, $P$ is the output laser power and $\omega$ is the radius of the Gaussian beam at a value of $1/\exp(2)$ of its maximum intensity.

Across each of the surfaces of the source, we consider the different energy loss mechanisms occurring during the heating of the source. The loss mechanisms transporting energy away from the laser irradiated spot within the source as a function of temperature, $T$, are compiled in the following list:\cite{materialscienceofthinfilms}
\begin{itemize}
	\item Conduction within the source ($\sim \nabla^2 T$)
	\item Radiation ($\sim \epsilon T^4$)
	\item Evaporation including overcoming the binding energy of the source material ($\sim {\exp(-1/T)}/{\sqrt{T}}$) and the kinetic energy distribution of the evaporant ($\sim \sqrt{T}\exp(-1/T)$).
\end{itemize}
In order to simplify the model, we disregard higher order energy loss mechanisms such as convection of molten material.
Due to the design of the free standing sources, thermal contact between the cylindrical source of material and the source holder itself is reduced, minimizing conduction across the boundaries, and therefore also disregarded.\cite{braun,smart} 

In addition to these loss mechanisms, the attenuation of the laser by the flux of evaporating material affects the incident energy flux upon the source. This can be modelled via an exponential decay ($\sim \exp(-\tau)$) in relation to a so-called optical depth, $\tau$.

Furthermore, we simplify the modelling of the evaporation process by assuming that the required energy to remove a particle from the condensed phase is characterized by the temperature-independent enthalpy of vaporization, $\Delta H_{\text{vap}}$.
Strictly speaking, this quantity should depend on $T$ due to the changing internal energy of the condensed phase, although there is little agreement regarding the most appropriate expression.
This fact is further exacerbated by the lack of experimental data for many materials.\cite{ALIBAKHSHI201762,Sipkens2018,Fish1975}
The approach of assuming a temperature-independent enthalpy of vaporization is consistent with other models of CW laser heating.\cite{Gallais2021}

We split the surface domain, $\p\Omega$, into two distinct (non-overlapping) regions: \lq Surface\rq\ $\partial\Omega_\text{S}$, and \lq Top Surface\rq\ $\partial\Omega_\text{TS}$.
For the \lq Surface\rq\ boundary regions, energy is lost from the outer surface of the source, namely via radiation and evaporation.
The \lq Top Surface\rq\ region additionally includes the incident heat flux on the surface from the laser.
On these surfaces with the surface normal, $n_i$, facing outwards from the body, we model:
\begin{widetext}
\begeq
- q_i n_i
=
\begin{cases}
& \displaystyle \epsilon\sigma (T^4 - T_{\text{amb}}^4) 
+ \bigg(\displaystyle \frac{\Delta H_{\text{vap}}}{M} + \frac{3N_\text{A}k_b T}{2M}\bigg)\sqrt{\frac{m}{2\pi k_b T}}p(T) 
\ \  \forall \ten x \in \partial\Omega_\text{S}
\\
& \displaystyle  - L(x,y)\exp(-\tau) + \epsilon \sigma (T^4 - T_{\text{amb}}^4) 
+ \bigg(\frac{\Delta H_{\text{vap}}}{M} + \frac{3N_\text{A}k_b T}{2M}\bigg)\sqrt{\frac{m}{2\pi k_b T}}p(T)  
\ \  \forall \ten{x} \in \partial\Omega_{\text{TS}}
\end{cases}
\label{surfaceequation}
\eqend
\end{widetext}
with parameters defined as follows:
\begin{itemize}
	\item $N_\text{A}$ - Avogadro's constant (1/mol)
	\item $\kappa$ - thermal conductivity (W/m\,K)
	\item $T_{\text{amb}}$ - ambient temperature of the chamber (assumed to be 300\,K)
	\item  $\epsilon$ - total hemispherical emissivity
	\item $\sigma$ - Stefan--Boltzmann constant (W/m$^2$K$^4$)
	\item k$_b$ - Boltzmann constant (J/K)
	\item $\Delta H_{\text{vap}}$ - enthalpy of vaporization of element (J/mol)
	\item $R$ - ideal gas constant (J/mol\,K)
	\item $m$ -  mass of evaporated particle (assumed to be the atomic mass of the source material) (kg)
	\item $M$ - molar mass of element (kg/mol)
	\item $p$ - vapor pressure (Pa)
\end{itemize} 
This system of equations is nonlinear due to the boundary terms.
We therefore solve it approximately by numerical analysis.
We employ the Finite Element Method (FEM) utilizing the property of monotonous convergence such that \textit{a posteriori} error analysis calculates the accuracy of the numerical approximate solution. In simple terms, the simulation is reliable when all prior assumptions are realistic. 

\begin{figure}[!b]
	\centering
	\includegraphics[width =\linewidth]{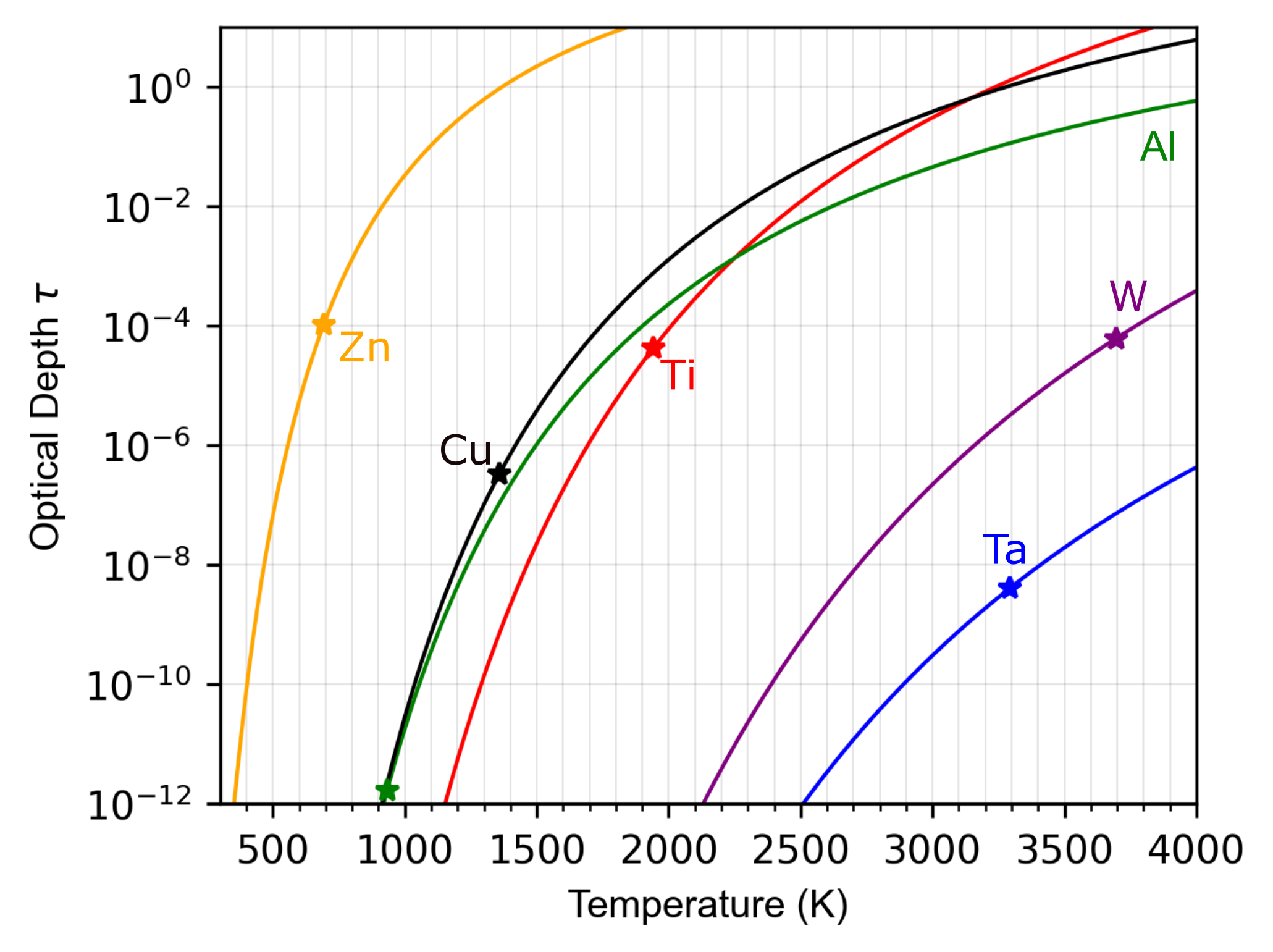}
	\caption{Optical depth $\tau$ for a range of elemental vapors for a laser wavelength $\lambda = 1030$\,nm as a function of temperature $T$ calculated using the expression in Eqn. \eqref{taumodel}.
	The stars indicate the melting point of the indicated element.
	Data regarding the atomic transitions of each element was obtained from the NIST database.\cite{NIST_ASD}}
	\label{Fig2}
\end{figure}

The attenuation of the laser beam by the evaporating species is quantified by the optical depth $\tau$, which is estimated from a modified version of the Beer--Lambert law:\cite{BLLaw}
\begin{equation}
\tau = \int n \sigma \dd r,
\end{equation}
where $n$ is the number density of the evaporant and $\sigma$ is the interaction cross section.
The interaction cross section accounts for all interactions between the laser photons and the elemental vapor, primarily absorption and scattering.
For the dilute vapor that is being considered here, the scattering cross section is significantly smaller than the absorption cross section and can thus be neglected.\cite{Foot2005}
In the case where the medium is homogeneous, the integral over distance $r$ gives the path length of the laser beam $l$.
In this case, we assume that the mass density of the evaporated material $\rho(r)$ follows an inverse-square law with $r$ from the $1/\exp(1)$ radius of the laser beam $\omega$ to the entrance port of the laser beam into the chamber $r_0$.
For the chamber geometry in which these experiments were performed, $r_0$ was equal to 500\,mm and the laser beam was incident upon the source at an angle of 45$^\circ$.
We can take the prior requirements and rewrite $n$ in terms of $\rho(r)$ to produce a new expression for $\tau$, as follows:
\begin{equation}
\tau = \int_{\omega}^{r_0} \frac{N_a \sigma \rho(r)}{M} \dd r, \rho(r) \sim \frac{1}{r^2}.
\end{equation}
Determining a general expression for $\sigma$ is challenging due to the variation of optical properties of atomic vapors across the periodic table.\cite{NIST_ASD}
Assuming that the transition is on resonance with the energy of the laser beam, we calculate an upper limit\cite{Foot2005} for the absorption cross-section: 
\begin{equation}
    \sigma_0 = \frac{3 \lambda_0^2}{2 \pi},
\end{equation}
where $\lambda_0$ is the resonant wavelength.
The true value of the absorption cross-section will be modified by the fact that it is unlikely that the laser wavelength coincides with a laser wavelength of a transition within the atomic species of the evaporant.
Therefore, we must account for the detuning of the laser relative to the resonant wavelength of the desired transition $\Delta$ and compare this to the rate of spontaneous emission\cite{Foot2005,Pizzey_2022}  from the excited state $\Gamma$,
\begin{equation}
\sigma = \sigma_0 \frac{\Gamma^2}{\Gamma^2 + 4\Delta^2}.
\end{equation}
For many metals, detuning rapidly reduces the value of $\sigma$, significantly affecting the optical depth of an atomic vapor.
In the case of Na, laser light of wavelength $\lambda = 600\,$nm gives $\sigma = 10^{-6}\sigma_0$ for the Na transition at $\lambda_0 = 589\,$nm.\cite{Foot2005}
Due to the variation in the onset of interband transitions between different elements, the value of the absorption cross-section varies significantly between the elements for the laser wavelength $\lambda = 1030\,$nm.
The optical properties of the vapor phase of many metallic gases have not been studied in detail; therefore, the optical depth must be treated as an estimate.
We emphasize that the expected power range where attenuation may become significant for many materials is beyond the maximum power of the laser used for these experiments.
Under the assumption that the gas is ideal---a valid assumption given the low pressure and high temperature of the atomic vapor\cite{YunusBook}---we then rewrite $\rho(r)$ in terms of the vapor pressure for the evaporant $p(T)$,\cite{vapour} for obtaining a final form:
\begin{equation}
\tau = \frac{3 \lambda_0^2}{2\sqrt{2} \pi k_b T} \frac{\Gamma^2}{\Gamma^2 + 4\Delta^2} \int_{\omega}^{r_0} \frac{p(T)}{r^2} \dd r.
\label{taumodel}
\end{equation}
For the case of Ta at $\lambda = 1030$\,nm, the closest absorption line is at 859\,nm,\cite{NIST_ASD} resulting in a detuning factor $\Delta$ of $\sim 10^{13}$\,Hz.
The lifespan of most excited electronic states in Ta are on the order of tens of ns, resulting in a value of $\Gamma \sim 10^8$\,Hz, making it negligible compared to $\Delta$.
For the melting point of Ta (3293\,K), $\tau \sim$ 10$^{-8}$, making it practically irrelevant for the power values considered within these experiments.
However, for Ti, interband transitions lie far closer to the laser wavelength of $\lambda = 1030$\,nm, with the nearest transition lying at $\lambda = 1032$\,nm, resulting in $\Delta \sim 10^{11}$\,Hz.
The calculated values of $\tau$ for a range of elements using the expression given in Eqn. \eqref{taumodel} as a function of $T$ are shown in Fig. \ref{Fig2}. This basic model of the optical depth does not fully consider the relative intensities of the closest transition within the relevant absorption spectra, which may impact the attenuation of the laser further. While $\tau$ is not significant in the power ranges considered within these experiments, it should be considered to get a full model of the laser heating process and the interaction between the incident laser and the evaporated material.

\subsection{Implementation and simulation}
\begin{figure}[!b]
	\centering
	\includegraphics[width =\linewidth]{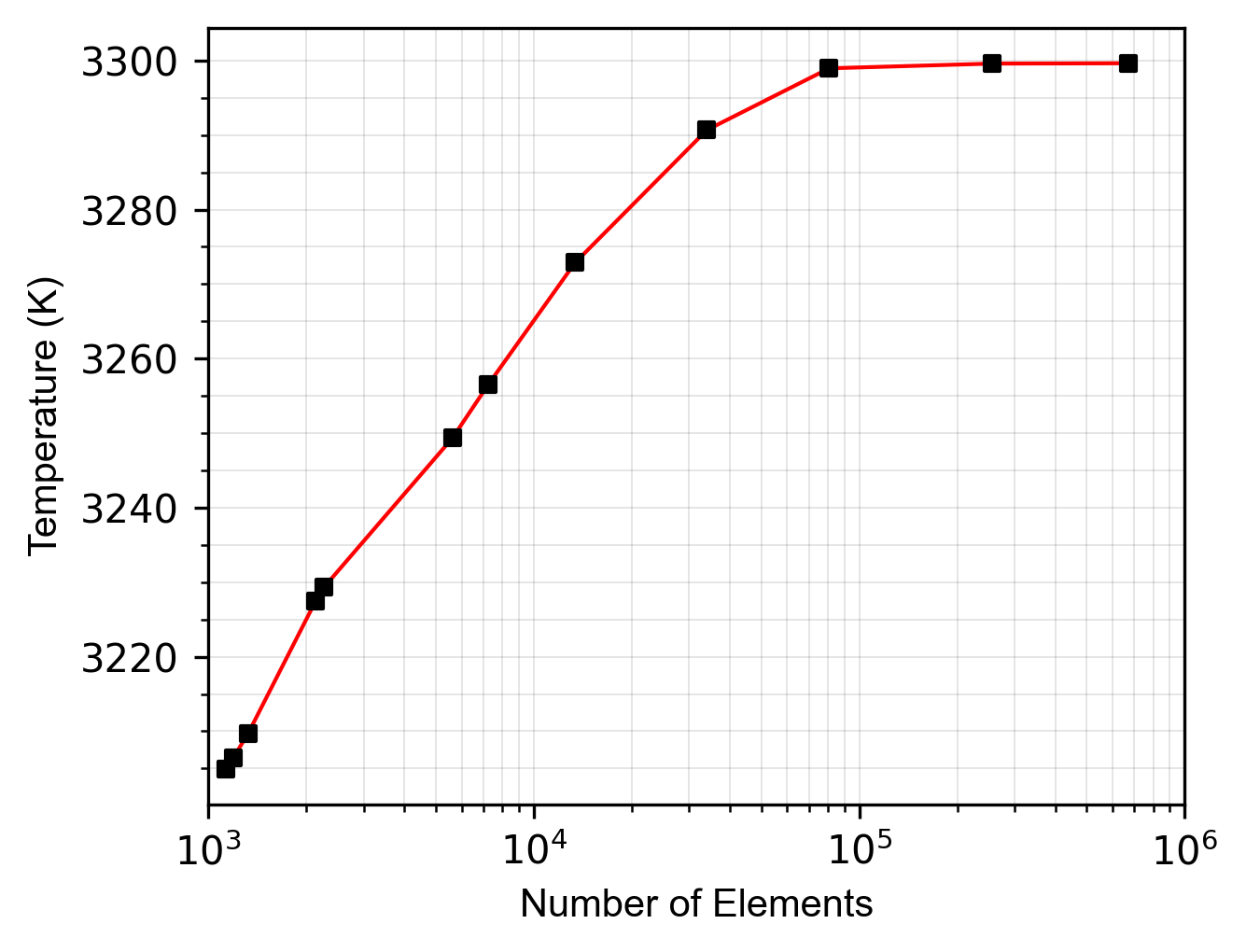}
	\caption{Graph showing how the maximum steady-state source temperature at the center of the laser spot converges to a solution with an increasing number of tetrahedral elements in FEM.
	This simulation was performed for a cylindrical source with a diameter of 3\,mm and a length of 8\,mm with $P$ = 280\,W and the laser spot centered on one of the end faces.}
	\label{Fig3}
\end{figure}

To construct the required form, we use a variational formulation and bring the governing equations, Eqn. \eqref{energybalance} and Eqn. \eqref{surfaceequation} into a weak form.
For this formulation, we discretize in time $T_0=T(t-\Delta t)$, by using the finite difference method with a constant time step $\Delta t$.
Therefore, the rate of temperature reads $(T-T_0)/\Delta t$, often called backward Euler method.
We use a discretization in space by FEM.
For the sake of simplicity, we disregard the distinction between analytical fields and their FEM representations as they never appear in the same line.
The representation over a finite number of nodal points (nodes) is employed for the unknown temperature $T$, by calculating at nodes and interpolating between them by means of basis functions.
We discretize the system by using (tetrahedron) Lagrange elements, which generates piece-wise continuous polynomials adequate for approximation in Hilbertian-Sobolev space $\mathcal{H}^1$.
These standard FEM elements\cite{zohdi2018finite,TheoryandPracticeofFE} of order $q$ span $\mathcal{P}_q$ on elements, $\Omega^\text{E}$ in a three-dimensional continuum.
In the computational domain, $\Omega$ is discretized by dividing it in elements (tetrahedrons), $\Omega^\text{E}$.
This so-called triangulation is denoted by $\mathcal{T}$.
Therefore, we use a function space, $\mathcal{V}$ for temperature with linear, $q=1$ elements:
\begin{equation}
\mathcal{V}=\bigg\{ 
		\big\{T\big\} \in \mathcal{H}^1(\Omega): \big\{T\big\}\Big|_{\Omega^\text{E}} \in \mathcal{P}_1(\Omega^\text{E}) \ \forall \Omega^\text{E} \in \mathcal{T} \bigg\} .
\end{equation}
As given by the Galerkin procedure, we utilize the same space for test functions of temperature $\del T$.
We use integration by parts for the term with second space derivatives and obtain the following weak form:
\begin{widetext}
\begal\label{weakform}
\text{Form} = &
\sum_\text{E}\int_{\Omega^\text{E}} \bigg( \rho c \frac{(T-T_0)}{\Delta t} \del T + q_i \del T_{,i} \bigg) \dd v 
- \int_{\p\Omega} q_i n_i \dd a 
\\
=& \sum_\text{E}\int_{\Omega^\text{E}} \bigg( \rho c \frac{(T-T_0)}{\Delta t} \del T - \kappa T_{,i} \del T_{,i} \bigg) \dd v 
+ \int_{\p\Omega_\text{S}} \bigg[\epsilon \sigma (T^4 - T_{\text{amb}}^4) 
+ \bigg(\frac{\Delta H_{\text{vap}}}{M} + \frac{3N_\text{A}k_b T}{2M}\bigg)\sqrt{\frac{m}{2\pi k_b T}}p \bigg] \del T \dd a 
\\ & 
+ \int_{\p\Omega_\text{TS}} \bigg[\bigg(\frac{\Delta H_{\text{vap}}}{M} + \frac{3N_\text{A}k_b T}{2M}\bigg)\sqrt{\frac{m}{2\pi k_b T}}p 
- \frac{(1-\mathcal{R})P \exp(-\tau) }{\pi w^2} \exp(-\frac{(x^2+y^2)}{w^2} ) \bigg] \del T \dd a \ .
\alend
\end{widetext}
The construction of the domain and discretization is conducted in Salome\cite{salome} by using NetGen algorithms.\cite{Schoberl1997}
Convergence tests are performed to observe the optimal number of tetrahedral elements required for the FEM simulation to converge to a steady state solution.
The results of these experiments are shown in Fig. \ref{Fig3} for a mesh with a diameter of 3\,mm and $P$ = 280\,W.
From these results, we conclude that approximately 10$^5$ tetrahedral elements are required for achieving optimal convergence whilst minimizing computation time.
An example of the mesh used for calculation is shown in Fig.~\ref{Fig1}.

For implementation, assembly, and solving, FEniCS packages are used within Python.
FEniCS is an open-source computing platform designed to solve partial differential equations.\cite{Fenics,LoggMardalEtAl2012,abali}
Since the weak form is nonlinear, linearization is employed by a standard Newton-Raphson solver and the incremental change of the solution is stopped by controlling the residuum in Eq.\,\eqref{weakform} since this form is equal to zero for the correct solution.
The transient solution of the above problem is run until it reaches a steady-state condition where the heat losses leaving the source equal the incoming heat flux from the laser, again within a predefined interval.

\section{Model Parameters}
The model has the following relevant thermophysical parameters that affect the result: $\kappa$, $c$, $\rho$, $\epsilon$ and $\mathcal{R}$.

Previous CW laser heating models explicitly consider the temperature dependence of each thermophysical parameter to accurately model the transient CW laser heating.\cite{Gallais2021,Elhadj2014,Combis2012}
However, the temperature dependence of these parameters is lacking from the literature for various materials, or is only given for a limited temperature range.
This lack of data presents a challenge for making the model as general as possible to allow for the modeling of any desired source material for TLE.
For a refractory metal like Ta, temperature dependent values of $\epsilon$ and $\mathcal{R}$ beyond 3000\,K are lacking.
$\mathcal{R}$ also has a significant wavelength dependence, further limiting the available temperature dependent data for the reflectivity of a material at the wavelength of the incident source laser.
Note that in thermal equilibrium for opaque bodies, $\epsilon$ and $\mathcal{R}$ should add up to unity.
The laser source, however, is a non-equilibrium narrow-band source.
Whereas $\epsilon$ is integrated over the entire black body spectrum, $\mathcal{R}$ is the $\mathcal{R}$ at 1030\,nm.

Equation \eqref{surfaceequation} has a substantial number of parameters, many of which are experimentally not well known.
We would like to know which of these dominate the behavior of the model, and which ones are less important.
Like this, we can develop a strategy to confirm the validity of the model.
In order to study the effect of each parameter on the steady state temperature, we therefore simulate the CW laser heating of a model Ta source with a diameter of 3\,mm and a length of 8\,mm.
In each simulation, we keep all other parameters constant and vary the parameter of interest across a large interval of hypothetical values.

\begin{figure}[!b]
	\centering
	\includegraphics[width =\linewidth]{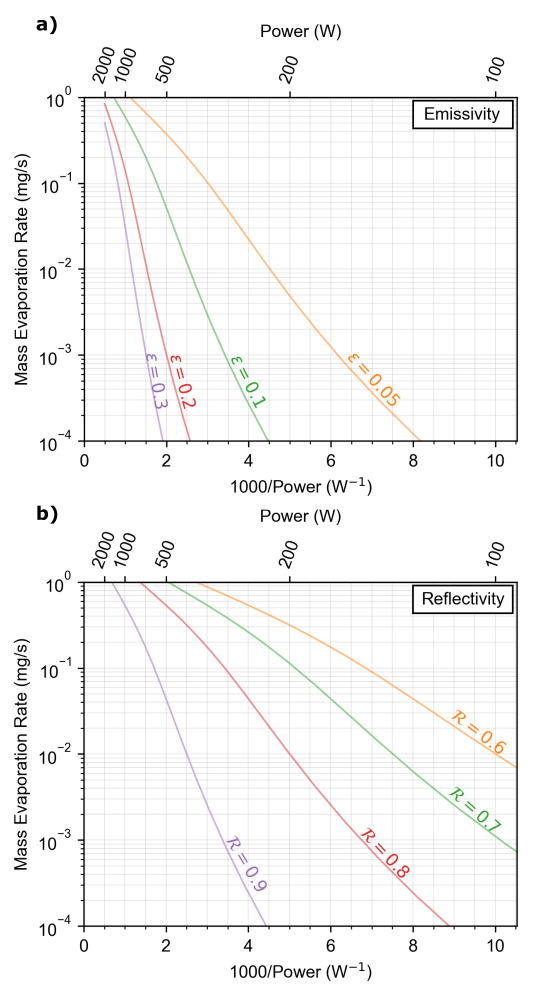}
	\caption{Panel a) shows the simulated mass evaporation rate of a Ta source with 3\,mm diameter and 8\,mm length for various values of the emissivity $\epsilon$.
	$\mathcal{R}$ was fixed at 0.87.
	Panel b) shows the simulated evaporation rate of the same Ta source for various values of the reflectivity $\mathcal{R}$.
	$\epsilon$ was fixed at 0.07.
	The other parameters were fixed at $\kappa$ = 57.5\,W/m\,K, $\rho$ = 16600\,kg/m$^3$, $c$ = 140\,J/kg\,K.}
	\label{Fig4}
\end{figure}

The result of this is shown in Fig.~\ref{Fig4} for $\epsilon$ and $\mathcal{R}$.
We clearly see that the simulated evaporation rate of Ta strongly depends on these parameters, giving mass evaporation rates that differ by more than three orders of magnitude over the ranges shown in Fig.~\ref{Fig4}.
Unfortunately, the range of values for these parameters in the literature for Ta are sparse and scatter over about the same range, see the green rectangle in Fig.~\ref{Fig7}, and therefore fail to effectively predict experimental results without additional reference data.

\begin{figure}[!b]
	\centering
	\includegraphics[width =\linewidth]{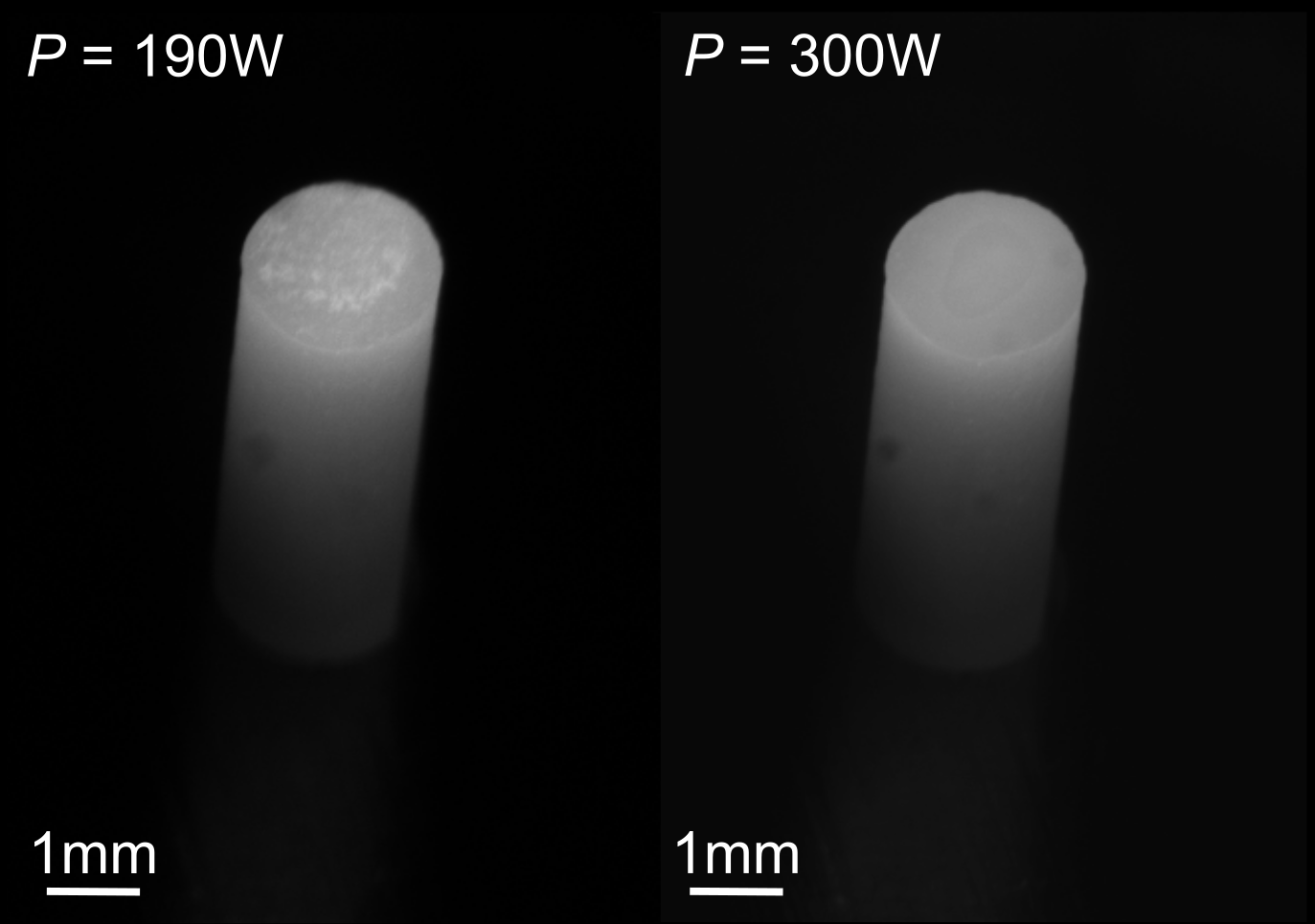}
	\caption{Photographs of a Ta source with a diameter of 3\,mm and a length of 8\,mm during irradiation with an infra-red $\lambda = 1030$\,nm laser beam with $\omega$ = 750\,\textmu m.
	The left panel shows that the irradiated surface of the Ta source remains solid at 190\,W while at 300\,W, a melt pool is visible at the location where the laser beam hits the surface.
	The melting point for the 3\,mm source of Ta was observed at 280\,W.}
	\label{Fig5}
\end{figure}

Fortunately, this reference information is readily available in TLE in the form of the melting point.
Under typical operating conditions for epitaxial deposition, most of the elements in the periodic table reach their melting points within the range of mass evaporation rates that are of technological interest.
We can therefore reliably connect a certain mass evaporation rate with the exactly know peak surface temperature on the source material when it starts melting.
Melting can be readily observed during operation with a video camera directed at the top surface of the source at the location of the peak laser intensity.

This is shown in Fig. \ref{Fig5} for a Ta source with a diameter of 3\,mm, where the melting point was observed at $P$ = 280\,W.
Once the melting point is reached, a localized liquid phase is observed on the surface of the Ta source. This observation links the laser power to the temperature at the laser spot for one specific temperature and the deposition rate, which can be measured, e.g.\ by determining the thickness of a deposited film under these conditions.

Out of these three parameters, the melting temperature can be determined with very high accuracy, given that the laser power is stable and can be varied by small enough amounts.
The mass evaporation rate can also be determined quite accurately, as \textit{ex-situ} film thickness measurements have a high resolution.
The laser power is less accurately known as it is typically measured in the laser, and possible attenuation in the delivery fiber, the beam shaping optics and the entrance window need to be taken into account.

In the following, we use this unambiguous reference point to investigate the contributions of the remaining parameters. We first pick reasonable values for $\epsilon$ and $\mathcal{R}$, such that the melting point temperature is reached at an incident laser power of 280\,W.
We then study the dependence of the source temperature when varying the remaining parameters $\kappa$, $\rho$, $\omega$ and $c$.
A strategy to more accurately determine $\epsilon$ and $\mathcal R$ will be discussed afterwards.

\begin{figure}[!b]
	\centering
	\includegraphics[width =\linewidth]{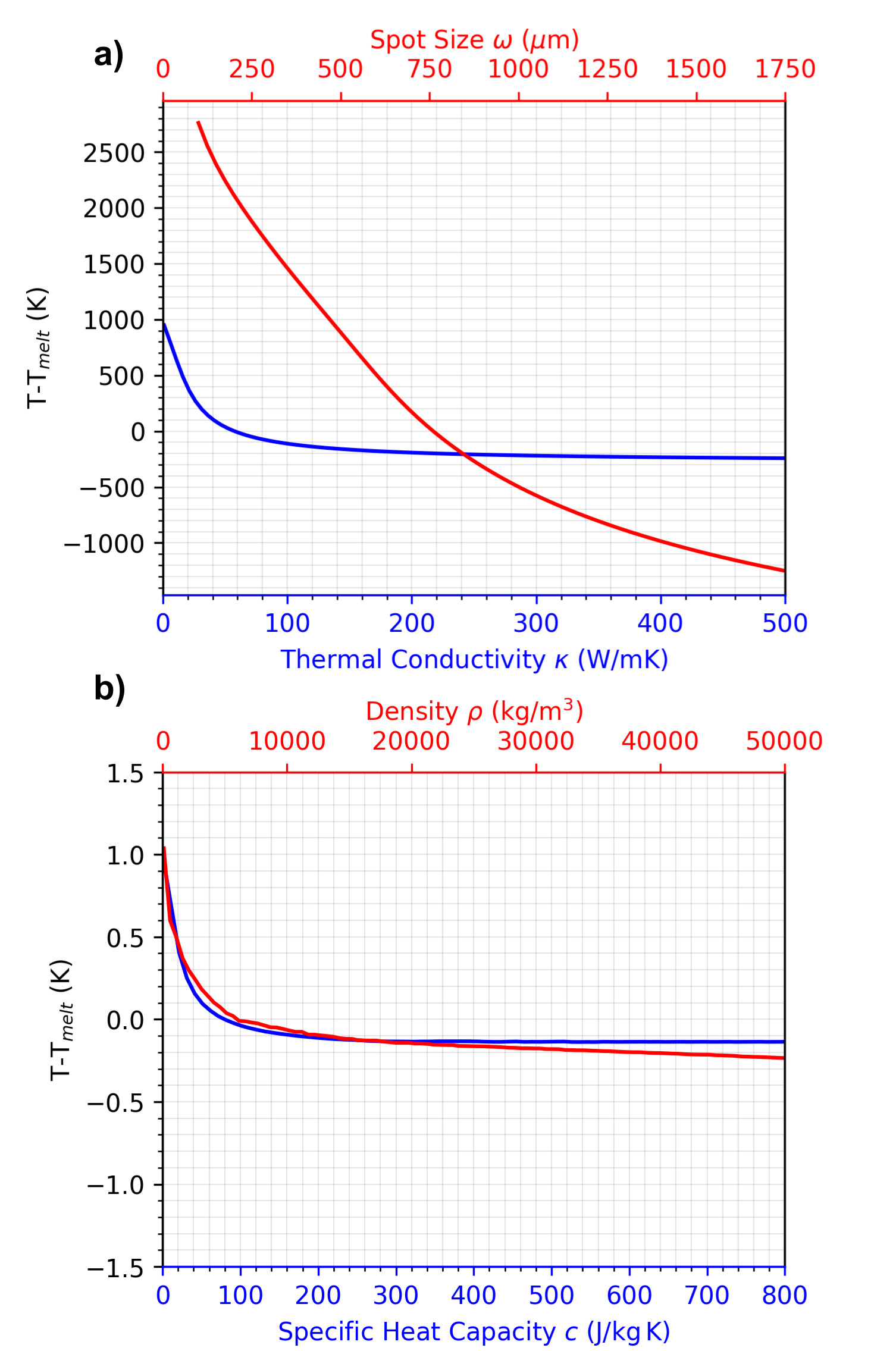}
	\caption{Variation in the peak source temperature relative to the melting point as output by finite element simulations of CW laser heating of a Ta source with a $\lambda = 1030$\,nm disk laser with $P$ = 280\,W.
	The source has a diameter of 3\,mm and a length of 8\,mm.
	Each panel shows the relative peak source temperature as one thermophysical parameter is varied whilst all others remain constant.
	Panel a) shows the effect of varying the thermal conductivity $\kappa$ and the spot size $\omega$ and Panel b) shows the corresponding effect for varying the specific heat capacity and the mass density of the source.
	For the corresponding simulations, the remaining physical parameters were as follows: $\mathcal{R}$ = 0.75, $\epsilon$ = 0.21, $\kappa$ = 57.5\,W/m\,K, $\rho$ = 16600\,kg/m$^3$ and $c$ =140\,J/kg\,K.}
	\label{Fig6}
\end{figure}
\begin{figure}[!b]
	\centering
	\includegraphics[width =\linewidth]{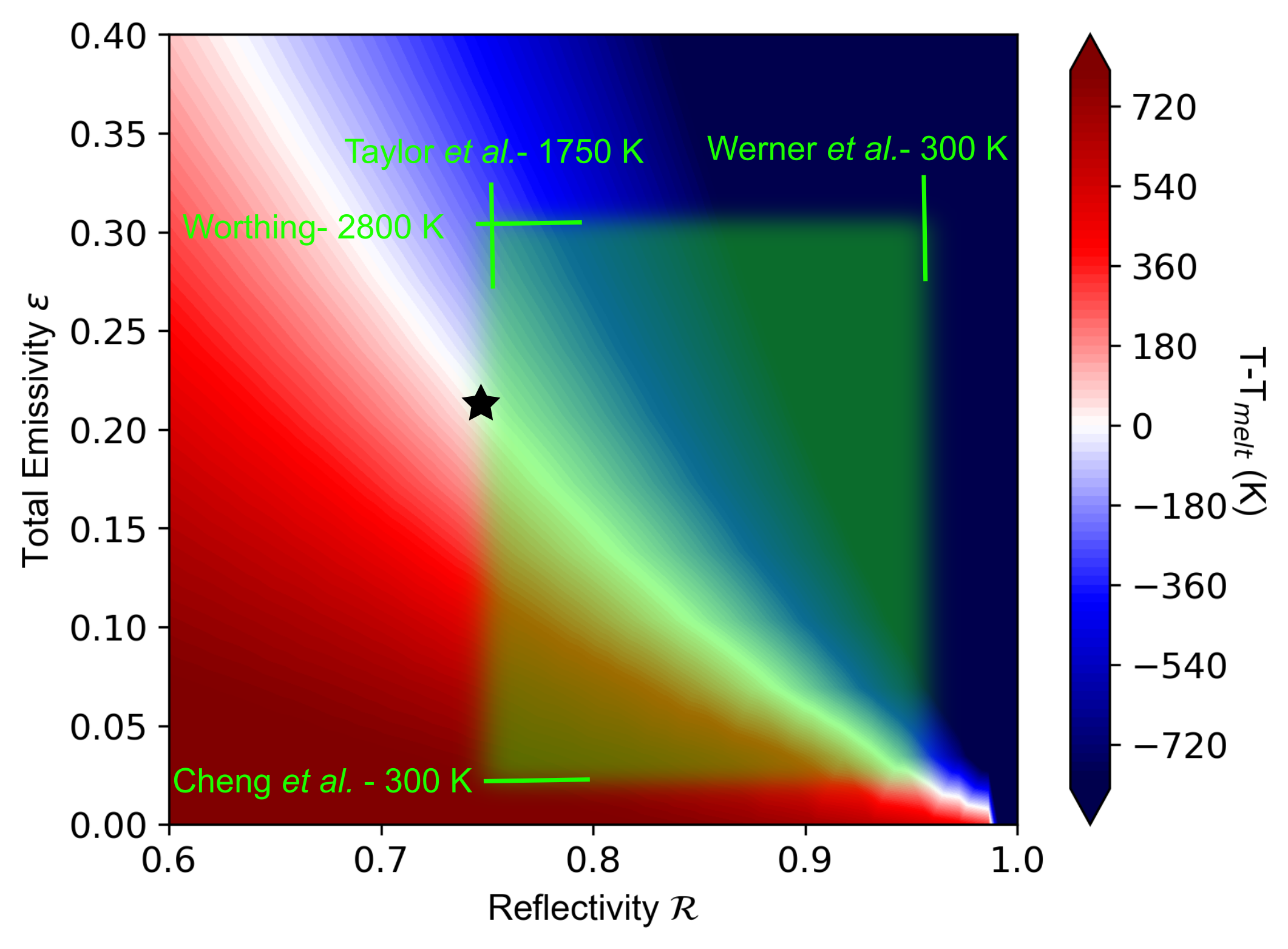}
	\caption{Parameter space showing the peak source temperature relative to the melting point of Ta obtained from finite element simulations for various values of $\epsilon$ and $\mathcal{R}$ with $P$ = 280\,W.
	The Ta source simulated had a diameter of 3\,mm.
	The green rectangle indicates the range of available temperature dependent values for $\epsilon$ and $\mathcal{R}$ in the literature for $\lambda = 1030$\,nm.\cite{Ordal1988,Cheng1987,Werner2009, Worthing1926,TaylorDewittreport}
	The selected $\epsilon$ and $\mathcal{R}$ pair found in best fit simulations is indicated by the black star.}
	\label{Fig7}
\end{figure}

Unlike the other parameters, one can calculate the spot size $\omega$ upon the source via geometrical optics using the dimensions of the TLE chamber and datasheet of the laser and its optics.\cite{Trumpf}
From this, we determine $\omega \sim 750\,$\textmu m for the given chamber geometry.
Uncertainty in the value of $\omega$ arises from the unknown tolerances of the focusing optics and possible distortions of the optical elements due to heating from absorbed laser light at these high powers.
The simulation is highly sensitive to the value of $\omega$ as demonstrated in Fig. \ref{Fig6}a.
This effect is more pronounced at low values of $\omega$ due to the intensity of the laser being inversely proportional to $\omega^2$ as introduced in Eqn. \eqref{lasersource}.
For the simulations shown in Fig. \ref{Fig6}, the incident laser power was fixed at 280\,W, the observed melting point of a Ta source with a diameter of 3\,mm.

We vary $\omega$ to test the effect of a small error in $\omega$ on the results of the model.
With a deviation of 10\,\% in $\omega$ at 750\,\textmu m, the error in temperature is 250\,K, which is large.
On the other hand, the error in $\omega$ is mostly systematic and therefore constant between the various experiments.
So it gets incorporated in the fit values for epsilon and R, yielding reliable predictions for TLE experiments.
However, in order to use TLE to obtain reliable values for $\epsilon$ and $\mathcal R$, the precise determination of omega is important. The output laser power can be assumed to be stable within 0.1\,\%.\cite{Trumpf}

The dramatic increase of the peak temperature, and therefore the evaporation or sublimation rate with smaller spot size $\omega$ indicates a corresponding increase in the efficiency of the TLE process for small laser spot sizes.
It is very much more efficient to move a small laser spot over the top surface of the source material, than to more or less uniformly heat the entire top surface with a large laser spot at the same laser power.

The variation of the thermal conductivity $\kappa$ is shown in Fig.~\ref{Fig6}a.
It is evident that varying $\kappa$ from low to high values initially has a significant effect on the maximum temperature of the source, with smaller values of $\kappa$ resulting in a highly localized temperature distribution with strong thermal gradients.
This is advantageous for efficient evaporation, since the mass evaporation rate is dominated by evaporation or sublimation from the hottest area, where a high temperature can be achieved with low laser power under these conditions.
As the value of $\kappa$ increases, thermal conduction becomes stronger, in the limit leading to a single uniform temperature for a given laser intensity.
This has the additional disadvantage that evaporation or sublimation takes place uniformly on all surfaces of the source material.
Under these conditions, it may therefore be useful to place the source material in a crucible which reflects and redirects these fluxes to the substrate.
Overall, however, the effect of $\kappa$ variation is less dramatic than the one of the spot size $\omega$.
And it is an immutable material parameter, which, unlike the laser spot size, is not available for process optimization with a given source material.

The variation of $\rho$ and $c$ is shown in Fig.~\ref{Fig6}b.
For time-independent simulations that search for the final steady state temperature of the source at the location of the laser spot, $\rho$ and $c$ have little effect (under 0.05\% variation for the range of values tested).
These primarily affect the transient temperature of the system as a function of time, once the piece of source material is charged with heat, the situation that is investigated here, their effect is negligible compared to the other parameters.
On the other hand, the values of $\rho$ and $c$ are significant for heating and cooling as these affect the required time for this system to reach a steady state, an important point in technological process optimization, where process throughput is critical.
The time dependence of this model can therefore be used to almost independently refine these parameters, which is convenient and increases the accuracy if their values need to be approximated from fits.
 
As discussed previously, there is a lack of temperature-dependent data for $\epsilon$ and $\mathcal{R}$ at the desired laser wavelength for many metals.
This is further compounded by the fact that both $\epsilon$ and $\mathcal{R}$ are affected by surface modifications and the purity of the analyzed material, resulting in a large variation in quoted values within the literature.
We can, however, use the mentioned reference point of the laser power at the melting point for a given source material to obtain realistic values for $\epsilon$ and $\mathcal{R}$.

Analogous to the synthetic variation of the other model parameters, we investigate a similar, now two-dimensional parameter space for $\epsilon$ and $\mathcal{R}$ as shown in Fig.~\ref{Fig7}.
It shows the temperature difference to the melting point of Ta predicted by the finite element model for the range of pairs of $\mathcal{R}$ along the horizontal, and $\epsilon$ along the vertical axis.
The range of values for $\epsilon$ and $\mathcal{R}$ for Ta available in the literature are indicated by the green rectangle in Fig. \ref{Fig7}.\cite{Ordal1988,Cheng1987,Werner2009, Worthing1926,TaylorDewittreport}

The white, neutral line in the plot shows pairs of $\epsilon$ and $\mathcal{R}$ values that are consistent with the observed melting point and the corresponding laser power.
Together with the other, less sensitive and better known parameters of the model, this white line now represents the value pairs that satisfy energy conservation through the framework of the finite element simulations.
This links $\epsilon$ and $\mathcal{R}$ in a similar way like Onsager relations,\cite{groot1984} which can be viewed as an extrapolation of Kirchhoff's law of thermal radiation.\cite{Kirchoff1860}.
Note that these do not directly apply here as the Onsager relations link total spectral and hemispherical emissivity and total spectral and hemispherical reflectivity, whereas we are considering the reflectivity at the single laser wavelength only.

We expect to obtain a pair of realistic values for for $\epsilon$ and $\mathcal{R}$ from the line segment where $T-T_\text{melt} = 0$ contained within the range of values for these parameters found in the literature, represented by the green, semi-transparent rectangle, likely at or just outside the high-temperature limits of these values.
And indeed, the black star indicates the pair used for the best fit in the following simulations: $\epsilon = 0.21$ and $\mathcal{R} = 0.75$.  

\begin{figure}[!b]
	\centering
	\includegraphics[width =\linewidth]{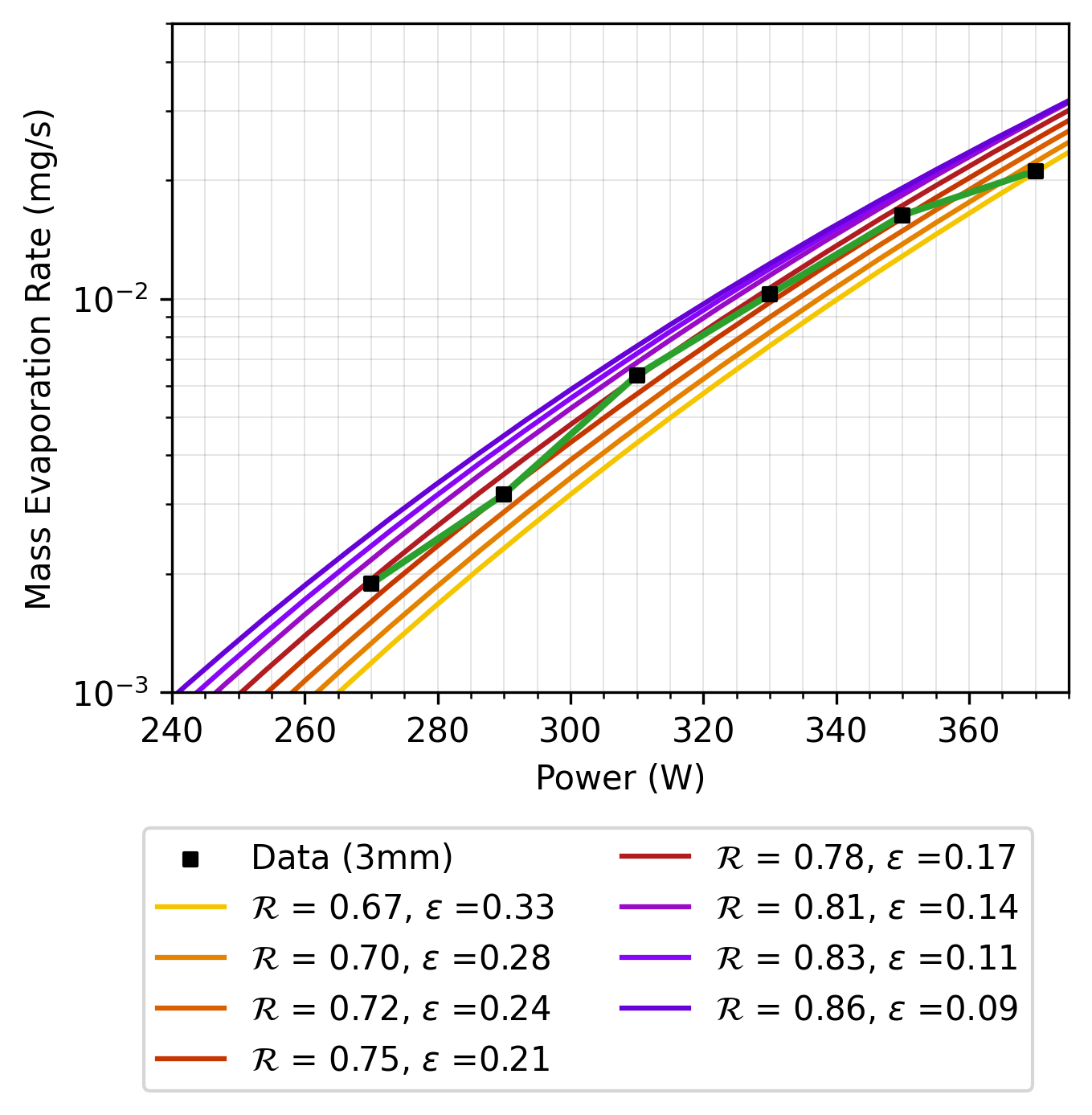}
	\caption{Effect of selecting different values along the line of $T-T_\text{melt}$ = 0 in Fig.~\ref{Fig7} on the mass evaporation rate for a Ta source with a diameter of 3\,mm.
	The best $\mathcal{R}$,$\epsilon$ pair may be determined by a fit to the experimental data. The experimental data was originally published in Ref \cite{SmartThesis}.}
	\label{Fig8}
\end{figure}

Selecting a differing pair of values along the line of $T-T_\text{melt}$ = 0 in Fig. \ref{Fig7} does affect the steady state result of the simulation.
For the temperature range of interest, we have identified two methods to narrow down the appropriate values of $\mathcal{R}$ and $\epsilon$ from the set of values provided by Fig. \ref{Fig7}.
The first concerns the steady-state result of the simulation for the mass evaporation rates.
In Fig. \ref{Fig8}, we plot the changes of the mass evaporation rates for a variation of $\mathcal{R}$ and $\epsilon$ along the white line of $T-T_\text{melt}$ = 0 in Fig.~\ref{Fig7} starting approximately from the horizontal center of the green box to beyond its left boundary, roughly centered around the black star.
The best fit of the various simulated curves to the experimental data points is obtained for $\epsilon = 0.21$ and $\mathcal{R} = 0.75$.

By repeating his procedure of finding meaningful values of $\epsilon$ and $\mathcal{R}$ via a parameter space then refining the ideal values by reducing residuals between the data and the model different source shapes and elements in the periodic table, one may oversample the available data space, perform consistency checks and iteratively refine the other parameters of the simulation.
Further such fits are presented in the experimental section below.

\begin{figure}[!b]
	\centering
	\includegraphics[width =\linewidth]{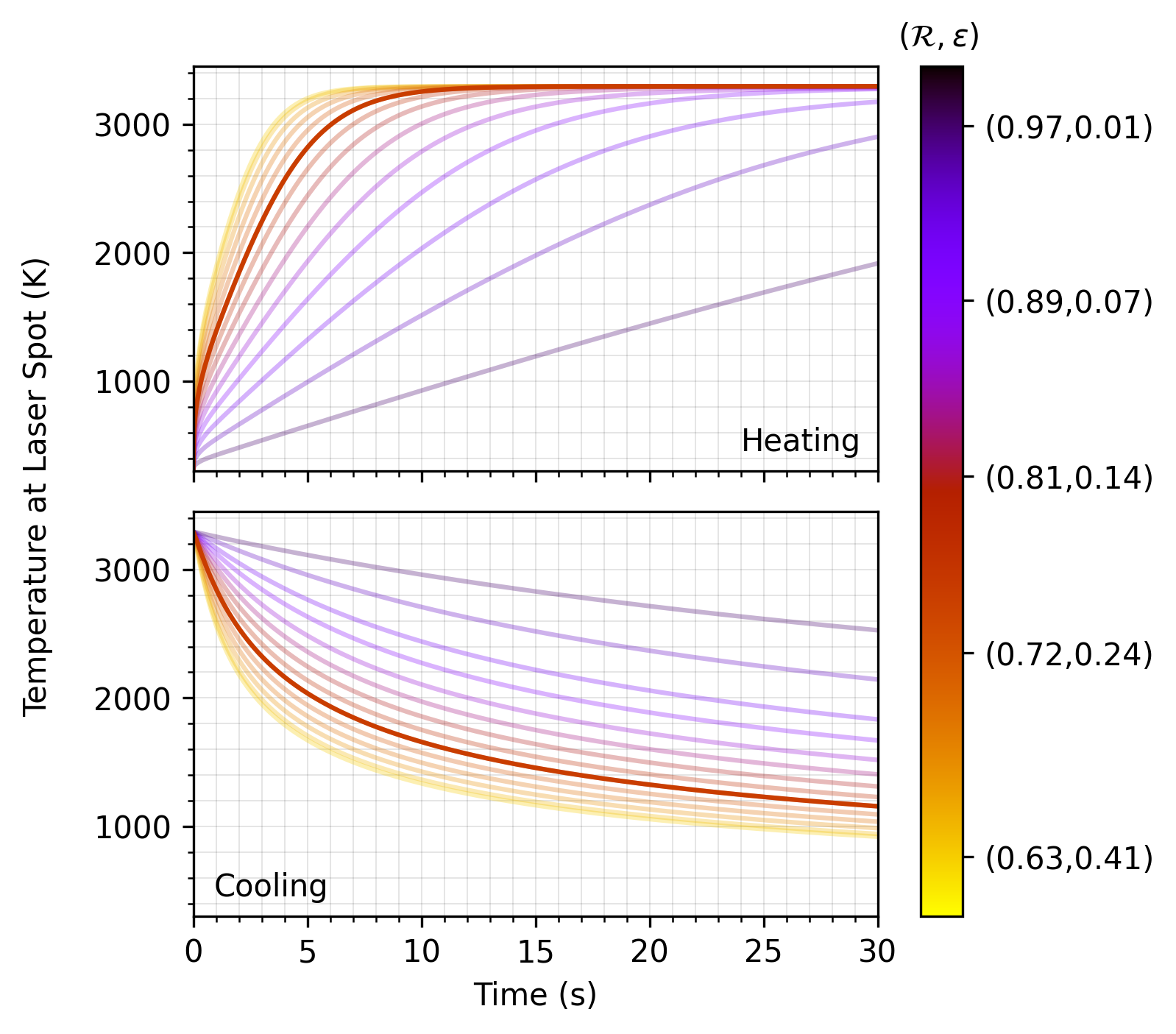}
	\caption{Simulated temperature transients during heating and cooling of a Ta source with a diameter of 3mm. The temperature was simulated at the center of the laser spot upon the source.
	The values of $\mathcal{R}$ and $\epsilon$ cover the range shown in Fig.~\ref{Fig7}.
	The other thermophysical constants follow commonly accepted temperature dependencies.\cite{Ta-cp,Ta-rho}
	An appropriate pair of values for $\mathcal{R}$ and $\epsilon$ may be determined from corresponding experiments once they become available, independently of the method shown in Fig.~\ref{Fig8}. The corresponding heating and cooling curves for $\epsilon = 0.21$ and $\mathcal{R} = 0.75$ are highlighted.
	}
	\label{Fig9}
\end{figure}

Another, largely independent approach to further refine the values of $\mathcal{R}$ and $\epsilon$ is via the temperature transients at the center of the laser spot during heating and cooling.
When turning the laser on or off using an abrupt step function, the temperature response of the source may be measured and compared to the simulation. 
Different values of $\mathcal{R}$ and $\epsilon$ strongly affect the rate of change of $T$ as a function of time.
This is illustrated in Fig. \ref{Fig9} for the case of $P$ = 280\,W and $\omega$ = 750\,$\mu$m, the previous set of parameters, corresponding to the melting point for a Ta source with a diameter of 3\,mm.
From these simulations, it is clear that the selection of $\mathcal{R}$ and $\epsilon$ has a significant effect on $T$ as a function of time and therefore can be fit to an experimental heating curve to help refine values of $\mathcal{R}$ and $\epsilon$.
Furthermore, the value of $\mathcal{R}$ is not relevant for the cooling curves and therefore the combination of both heating and cooling curves can be used to extract $\mathcal{R}$ unambiguously.
In addition, an independent verification or refinement of $\kappa$, $\rho$ and $c$ becomes possible with such measurements, as these now play a  significant role in the behavior of the system.

Note that heating and cooling takes place within seconds for such a TLE source, allowing its operation without a shutter, such that the flux is initiated and terminated by modulating the laser light intensity alone.

The experimental measurements of such curves remains a subject of further study.
Again, the melting point may serve as an accurate and reliable calibration point to the experimental data for such measurements as well.

\section{Experiments}
To obtain the experimental mass evaporation rate data for various Ta sources, we use a greatly simplified TLE chamber as shown in Fig. \ref{Fig1}a.
A single Ta source is locally heated inside of a vacuum chamber by a $\lambda = 1030$\,nm disk laser with a Gaussian beam shape and $\omega=750\,\mu$m.
The chamber is not actively cooled and operates at ultra high vacuum (UHV) pressures between 10$^{-8}$-10$^{-9}$\,mbar.
The substrate on which a Ta film is deposited is a 2" Si (100) wafer which was selected due to its low cost, purity and UHV compatibility.
The working distance between the source and substrate is 60\,mm.

All the sources inserted into the chamber are cylindrical and have heights of 8\,mm with various diameters. 

The average growth rate of Ta for a fixed incident laser power is measured using a quartz crystal microbalance (QCM) after a deposition time of 20 minutes.
These values are converted to mass evaporation rates by measuring the mass of the Ta source before and after deposition for a laser power value where significant evaporation occurs over the 20 minute deposition period.
The mass difference is then divided by the deposition time to obtain the average mass evaporation rate.
This rate produces a geometrical conversion factor to convert the growth rate data to mass evaporation rate data for each Ta source.
This factor only depends on the geometry of the evaporation system and thus remains the same for each Ta source and any laser power.
By scaling this factor against the ratio of mass density values for a given element compared to Ta, this conversion factor may be used to convert the growth rate data of any element.

\section{Results}

\begin{figure}[!b]
	\centering
	\includegraphics[width =\linewidth]{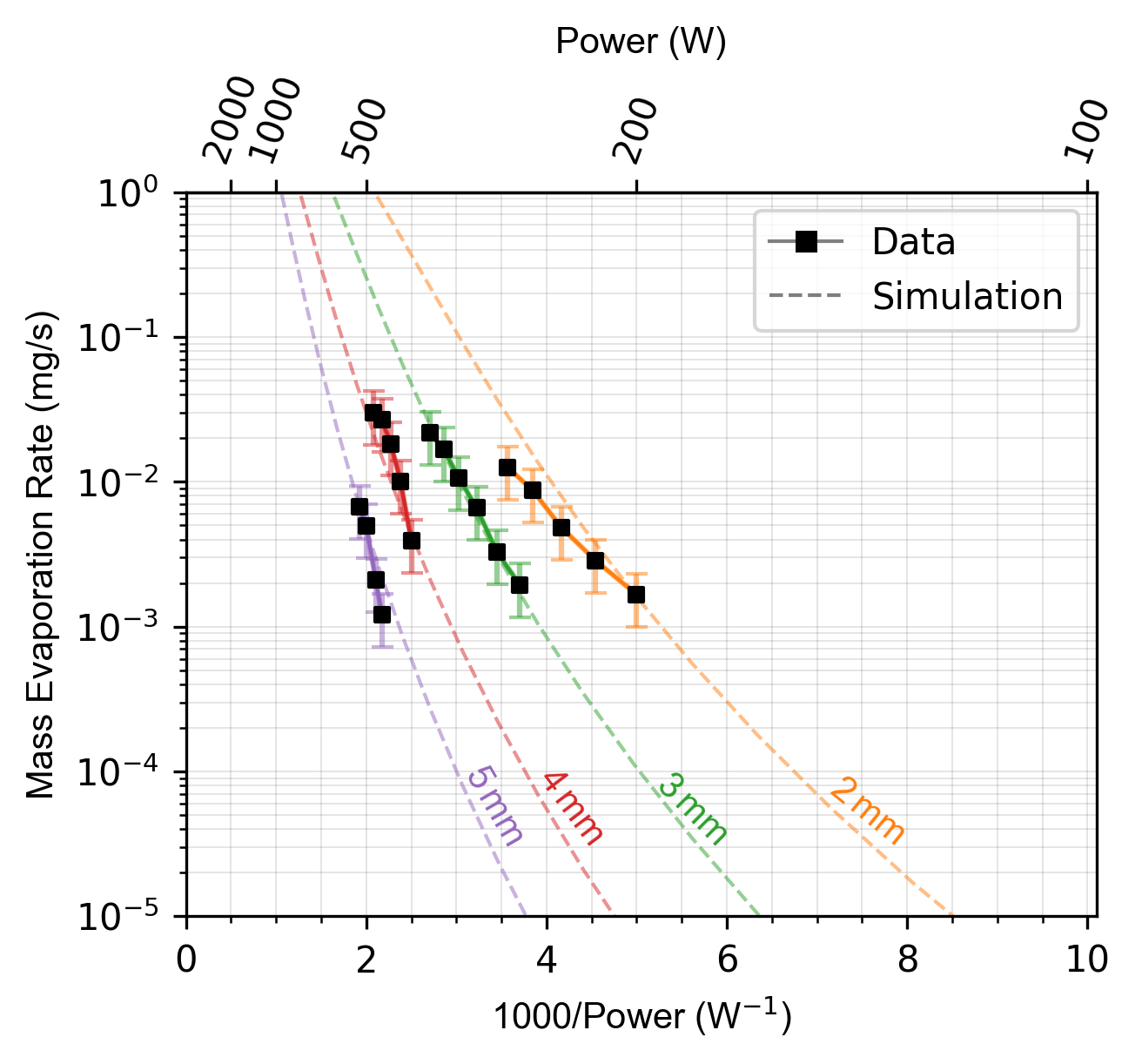}
	\caption{Comparison between experimental and simulated mass evaporation rate data of Ta sources with various diameters.
	Each experimental data point was scaled from growth rate data recorded by QCM over a sequence of 20 minute depositions at the corresponding laser power. The experimental data was originally published in Ref \cite{SmartThesis}.}
	\label{Fig10}
\end{figure}

We performed a comparison between the simulated and experimental mass evaporation rate data of Ta sources with various diameters as shown in Fig. \ref{Fig10}.
The experimental mass evaporation rate data was scaled from growth rate data obtained via the QCM, using an experimentally derived scaling factor (0.033 $\pm$ 0.006\,mg/\AA) as previously described.

The simulation simultaneously fits the data for a range of source diameters while keeping all other parameters constant, indicating that the model is valid and has predictive power.

\begin{figure}[!b]
	\centering
	\includegraphics[width =\linewidth]{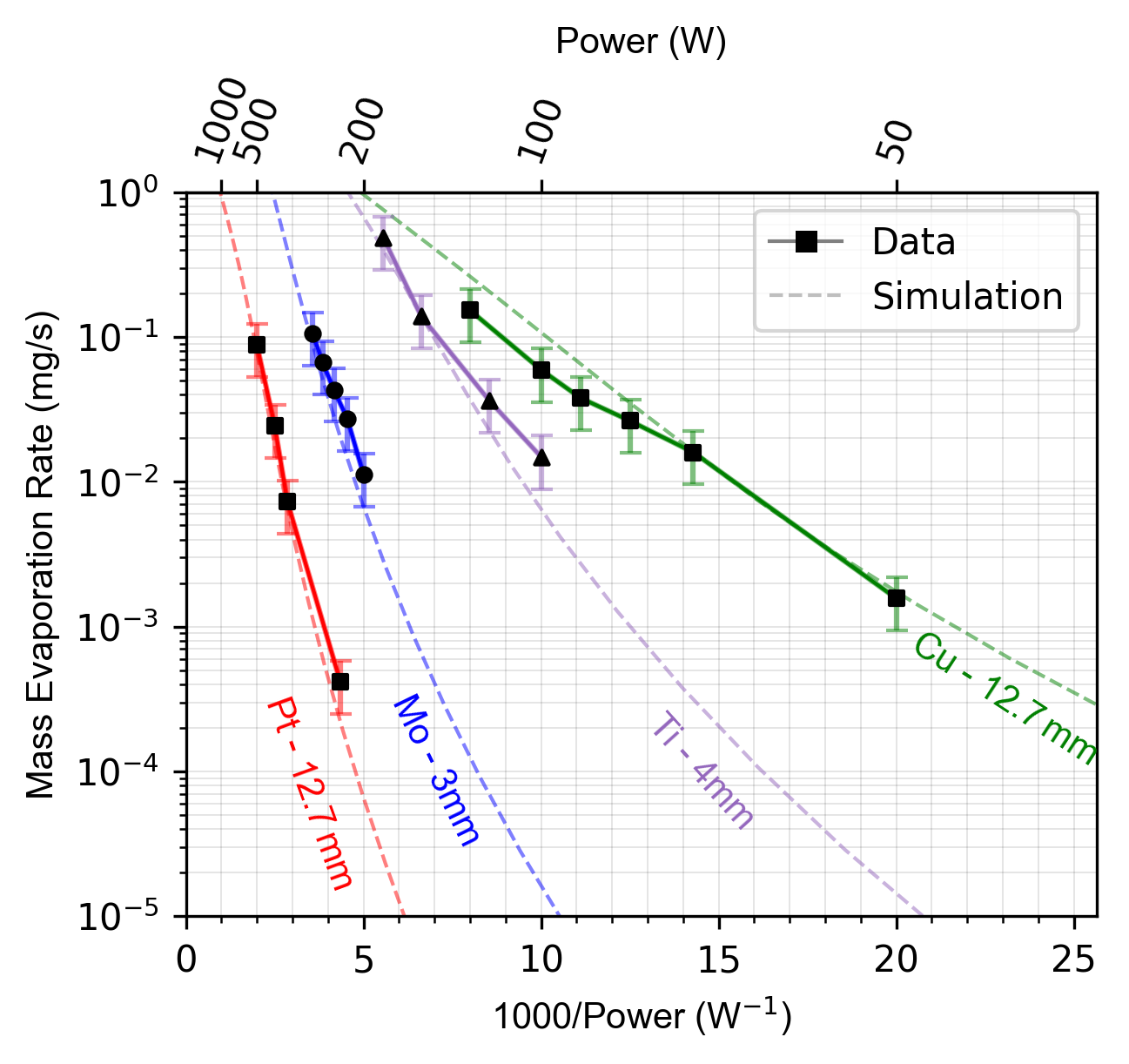}
	\caption{Comparison between experimental and simulated mass evaporation rate data of various sources with a range of compositions and diameters.
	The values of $\mathcal{R}$ and $\epsilon$ used for each material are given in Table~\ref{table:1}. The experimental data was originally published in Ref \cite{SmartThesis}.
	}
	\label{Fig11}
\end{figure}

We can now follow the same algorithm as for Ta and model the mass evaporation rates of other elements.
We have done this for four other materials: Pt, Mo, Ti, and Cu.
These materials are selected due to their wide range of physical properties, primarily the range of reflectivity values and vapor pressures.
All materials except Cu were illuminated with the same laser beam as the Ta source with $\lambda$ = 1030\,nm and $\omega$ = 750\,$\mu$m.
For Cu, we used a frequency-doubled $\lambda$ = 515\,nm with a peak power of 1\,kW and an identical $\omega$ = 750\,$\mu$m.
The results of the simulations of these materials are collected in Fig.~\ref{Fig11}.
Results of the calculated effective values of $\mathcal{R}$ and $\epsilon$ determined in the same way as discussed above for Ta are compiled in Table\,\ref{table:1}. 

\begin{table}[h!]
\caption{Table of calculated effective values of $\mathcal{R}$ and $\epsilon$ for various elements to obtain the evaporation rate data illustrated in Fig. \ref{Fig11}.}
\label{table:1}
\setlength{\tabcolsep}{0.8em}
\renewcommand{\arraystretch}{1.8}
\begin{tabular}{|c|r|r|r|r|} 
 \hline
 Element & Diameter (mm) & $\lambda$ (nm) & $\mathcal{R}$ & $\epsilon$  \\ [0.5ex] 
 \hline
 Pt & 12.7 & 1030 & 0.73 & 0.18 \\ 
 \hline
 Mo & 3.0 & 1030 & 0.66 & 0.36 \\
 \hline
 Ti & 4.0 & 515 & 0.67 & 0.55 \\
 \hline
 Cu & 12.7 & 515 & 0.64 & 0.23 \\ 
 \hline
\end{tabular}
\end{table}

\section{Discussion and Conclusion}
The results shown in Figs.~\ref{Fig10} and \ref{Fig11} demonstrate that our model can accurately describe CW laser heating at elevated temperatures.
This is possible despite the lack of independent temperature dependent data for the thermophysical properties of many metals at temperatures close to and beyond their corresponding melting points, as these may be determined self-consistently from calibrating the model to the reference provided by the required laser power to reach the melting point of a given source.

In order to effectively simulate the steady state evaporation of elemental sources in TLE, we find that the relevant physical parameters are $\kappa$, $\mathcal{R}$ and $\epsilon$, since they dominate the behavior in the regimes relevant for practical deposition applications.
Whilst extensive studies of the temperature dependence of $\kappa$ have been performed for most elements,\cite{Ta-kappa} corresponding data for $\mathcal{R}$ at $\lambda$ = 1030\,nm and the total emissivity $\epsilon$ are lacking for many elements.

One potential challenge is the investigation of the attenuation effect of the source laser by the evaporating material.
Experimental data at high laser powers must be obtained in order to investigate this and see how this compares with the model given in Eqn. \eqref{taumodel}.
Within the typical operating range of TLE, however, this is unlikely to be noticeable unless extremely high deposition rates were desired, or a transition of the vapor atoms or molecules is in resonance with the laser wavelength.

The method of parameter determination relies on the melting point of a material being observable in the experiment.
If a material does not melt in UHV conditions,for example, carbon, then this algorithm is unsuitable and requires independently determined temperature-dependent $\mathcal{R}$ and $\epsilon$ values to accurately simulate.

In addition, the incident source laser is assumed to be a surface source as the penetration depth of laser radiation at $\lambda$ = 1030\,nm and $\lambda$ = 515\,nm is on the order of nanometers for many metals.
Consequently, the construction of the incident source laser would need to be switched to a volumetric source for materials where the penetration depth is not negligible compared to the dimensions of the source.

The good agreement between the finite-element simulations and the experiments across a range of materials and geometries indicates that the procedure correctly models the main mechanisms involved in the process.
For practical applications in TLE, the motivation of this work, these simulations allow an accurate prediction of the process, and thereby accurate process control.
We expect that further refinement of the material parameters becomes possible as more experimental data of both static and time-dependent experiments becomes available.
Since laser evaporation of ultra-pure single elements in vacuum is a well defined experimental system, such experiments also provides a method to determine the fundamental properties of the elemental materials at extremely high temperatures.

\section{Acknowledgments}
The authors thank Peter Schüffelgen, Alexander Pawlis, Kristof Moors, Christine Falter and Laurent Gallais for many insightful scientific discussions. We are grateful to Fabian Felden for technical help with experimental measurements and to Dennis Heffels for technical assistance with simulations.

\section{Data Availability}
The data and code supporting the findings of this study are available from the authors upon reasonable request.
\section{References}
\bibliographystyle{unsrtnat}
\bibliography{fea}
\end{document}